\documentclass[12pt]{article}
\newcommand{\beq}{\begin{equation}}
\newcommand{\eeq}{\end{equation}}
\newcommand{\bra}{\begin{array}}
\newcommand{\era}{\end{array}}

\newcommand{\al}{\alpha}

\newcommand{\de}{\delta}

\newcommand{\ep}{\epsilon}
\usepackage{latexsym}
\author{Jamila Douari\footnote{jdouari@gmail.com}\\ \\
\small\it Center for Advanced Mathematical Sciences\\\small\it
American University of Beirut\\ \small\it P.O.Box 11-0236, College
Hall \\\small\it Beirut, Lebanon\rm}
\title{Electrified Higher-Dimensional Brane Intersections}

\begin{document}
\maketitle \vspace*{0.5cm} PACS: 11.25.-w, 11.25.Uv, 11.15.Kc
\vskip1cm Keywords: Intersecting
Branes, Dyons, Duality. \vspace*{1.5cm}
\section*{Abstract}
\hspace{.3in}We discuss the duality of intersecting D1-D5 branes in
the low energy effective theory in the presence of electric field.
This duality is found to be unbroken. We also deal with the
solutions corresponding to two and three excited scalars in the
D3-brane theory in the absence and the presence of an electric field.
The solutions are given as a spike which is interpreted as an
attached bundle of a superposition of coordinates of another brane
given as a collective coordinate a long which the brane extends away
from the D3-brane. The lowest energy in both cases is higher
than the energy found in the case of D1$\bot$D3 branes.
\section{Introduction}
\hspace{.3in}The Born-Infeld (BI) action \cite{BI} governs the
D-brane's world volume which has many fascinating features. Among
these there is the possibility for D-branes to morph into other
D-branes of different dimensions by exciting some of the scalar
fields \cite{InterBran1,InterBran2}. The subject of intersecting
branes in string theory is very rich and has been studied for a
long time \cite{inter}. One of the main sections of the present
paper will be devoted to discuss the duality of intersecting D1-D5
branes and another section studies the intersection of D2- and
D3-branes with a D3-brane.

One of the interesting subjects we discuss in this work
concerns the duality of D1$\bot$D5 branes in the presence of
electric field. By considering the energy of our system from D1 and
D5 branes descriptions we find that the energies obtained from the two
theories match. Consequently, the presence of an electric field doesn't spoil this duality as seen in D1$\bot$D3 branes case we showed in the reference \cite{brokdual}.

In the case of D1$\bot$D3 branes, it is known in the literature that
there are many different but physically equivalent descriptions of
how a D1 brane may end on a D3 brane. From the point of view of the
D3 brane the configuration is described by a monopole on its world
volume. From the point of view of the D1 brane the configuration is
described by the D1 opening up into a D3 brane where the extra two
dimensions form a fuzzy two sphere whose radius diverges at the
origin of the three-brane. These different view points are the
stringy realization of the Nahm transformation \cite{fun}.

By following the same mechanism used for D1$\bot$D3 branes, this
paper is devoted to finding the funnel solutions when two and three
excited scalars are involved. Thus, we restrict ourselves to using BPS
arguments to find some solutions which have the interpretation of a D2-brane ending on D3-brane
and D3-brane ending on other D3-brane. In these studies we consider
the absence and the presence of an electric field. We found that the
investigation of excited D3-brane in each case leads to
the fact that by exciting 2 and 3 of its transverse directions in
the absence or the presence of electric field, the brane develops a
spike which is interpreted as an attached bundle of a superposition
of coordinates of another brane given as a collective coordinate a
long which the brane extends away from the D3-brane. Then, by
considering the lowest energy states of our system we remark that the lowest
energy in the intersecting branes case is obtained by the
D1-D3 branes intersection and the energy is higher if we excite more
scalar fields and even more in the presence of an electric field.

This paper is organized as follows: In section 2, we review in brief
the funnel solutions of D1$\bot$D3 branes and its broken duality in the
dyonic case. In section 3, we discuss the unbroken duality of
intersecting D1-D5 branes in the presence of an electric field. In section 4, we investigate the two and three excited scalars in D3-brane theory and we conclude in section 5.
\section{D1$\bot$D3 Branes in Dyonic Case}
\hspace{.3in}In this section, we review in brief, the funnel
solutions for D1$\bot$D3 branes. First, we give the funnel solution
by using abelian Born-Infeld action for the world volume gauge field
and one excited transverse scalar in the dyonic case. It was shown in
\cite{cm} that the BI action, when taken as the fundamental action,
can be used to build a configuration with a semi-infinite
fundamental string ending on a D3-brane \cite{Gib}. The dyonic
system is given by using D-string world volume theory and the
fundamental strings are introduced by adding a $U(1)$ electric
field. The system is described by the following action \beq\bra{lll}
S=\int dt L &=-T_3 \int d^4\sigma\sqrt{-det(\eta_{ab}+\lambda^2
\partial_a \phi^i \partial_b \phi^i +\lambda F_{ab})}\\\\
&= -T_3 \int d^4\sigma\Big[ 1 +\lambda^2 \Big( \mid \nabla\phi
\mid^2 +\stackrel{\rightarrow}{B}^2 +\stackrel{\rightarrow}{E}^2
\Big)\\\\
&+\lambda^4 \Big( (\stackrel{\rightarrow}{B}.\nabla\phi )^2
+(\stackrel{\rightarrow}{E}.\stackrel{\rightarrow}{B})^2 +\mid
\stackrel{\rightarrow}{E}\wedge\nabla\phi\mid^2 \Big)
\Big]^{\frac{1}{2}}, \era\eeq in which $F_{ab}$ ($a,b=0,...,3$) is
the Maxwell field strength and the electric field is denoted by
$F_{0a}=E_a$. $\sigma^a$ denote the world volume coordinates while
$\phi^i$ ($i=4,...,9$) are the scalars describing transverse
fluctuations of the brane and $\lambda=2\pi \ell_s^2$ with $\ell_s$
is the string length. In our case we excite just one scalar so
$\phi^i=\phi^9 \equiv\phi$. The energy of the present dyonic system
accordingly to (1) is given by \beq\bra{ll}
E&= T_3 \int d^3\sigma\Big[ \lambda^2 \mid \nabla\phi
+\stackrel{\rightarrow}{B} +\stackrel{\rightarrow}{E}\mid^2
+(1-\lambda^2 \nabla\phi.\stackrel{\rightarrow}{B})^2-2\lambda^2
\stackrel{\rightarrow}{E}.(\stackrel{\rightarrow}{B}
+\nabla\phi)\\\\
&+ \lambda^4 \Big(
(\stackrel{\rightarrow}{E}.\stackrel{\rightarrow}{B})^2 +\mid
\stackrel{\rightarrow}{E} \Lambda\nabla\phi\mid^2\Big)
\Big]^{1/2}.\era\eeq Then if we require the condition that
$\nabla\phi +\stackrel{\rightarrow}{B}+\stackrel{\rightarrow}{E}=0$,
the energy $E$ reduces to the following positive energy \beq E_0
=T_3 \int d^3\sigma\Big[(1-\lambda^2
(\nabla\phi).\stackrel{\rightarrow}{B})^2+2\lambda^2
\stackrel{\rightarrow}{E}.\stackrel{\rightarrow}{E} + \lambda^4 (
(\stackrel{\rightarrow}{E}.\stackrel{\rightarrow}{B})^2
+{\mid\stackrel{\rightarrow}{E}}\Lambda \nabla\phi\mid^2)
\Big]^{1/2},\eeq as a minimum energy of the system. Thus, by solving
the proposed condition $\nabla\phi
+\stackrel{\rightarrow}{B}+\stackrel{\rightarrow}{E}=0$ the funnel
solution is found to be \beq\phi =\frac{N_m +N_e}{2r},\eeq
with $N_m$ is magnetic charge and $N_e$ electric charge.

Now, we consider the dual description of the D3$\bot$D1 system. In
D-string theory, we use non-abelian BI action given by the following
to get a D3-brane from $N$ D-string \beq S=-T_1\int d^2\sigma STr \Big[
-det(\eta_{ab}+\lambda^2 \partial_a \phi^i Q_{ij}^{-1}\partial_b
\phi^j )det Q^{ij}\Big]^{1\over2}\eeq
where $ Q_{ij}=\de_{ij} +i\lambda \lbrack \phi_i , \phi_j \rbrack$, $i=1,2,3$ and $a,b=\tau,\sigma$.

Expanding this action to leading order in $\lambda$ yields the usual
nonabelian scalar action
$$S\cong -T_1\int d^2\sigma  \Big[ N+ \lambda^2 Tr (\partial_a
\phi^i + \frac{1}{2}\lbrack \phi_i , \phi_j \rbrack \lbrack \phi_j ,
\phi_i \rbrack) +...\Big]^{1\over2}.$$
The solution of the equation of motion of the scalar fields $\phi_i$
represent the $N$ D-string expanding into a D3-brane analogous to the
bion solution of the D3-brane theory \cite{InterBran1,InterBran2}.
The solutions are
$$\bra{lc}\phi_i =\pm\frac{\al_i}{2\sigma},&\lbrack \al_i , \al_j \rbrack=2i\ep^{ijk}\al_k ,\era$$
with the corresponding geometry is a long funnel where the
cross-section at fixed $\sigma$ has the topology of a two-sphere.

The dyonic case is taken by considering ($N, N_f$)-string. We have
$N$ D-strings and $N_f$ fundamental strings \cite{9911136}. The
theory is described by the action \beq S=-T_1\int d^2\sigma STr
\Big[ -det(\eta_{ab}+\lambda^2 \partial_a \phi^i
Q_{ij}^{-1}\partial_b \phi^j +\lambda F_{ab})det
Q^{ij}\Big]^{1\over2}.\eeq This action can be rewritten as \beq
S=-T_1\int d^2\sigma STr \Big[ -det\pmatrix{\eta_{ab}+\lambda
F_{ab}& \lambda \partial_a \phi^j \cr -\lambda \partial_b \phi^i &
Q^{ij}\cr}\Big]^{1\over2}.\eeq The bound states of D-strings
and fundamental strings are made simply by introducing a background
$U(1)$ electric field on the D-strings corresponding to fundamental
strings dissolved on the world sheet. Then we replace the field strength $F_{\tau\sigma}$ by $EI_{N_m}$ ($I_{N_m}$ is
$N_m \times N_m$-matrix) and the following ansatz is inserted
\beq\phi_i =\hat{R}\al_i .\eeq
For non-abelian scalars and at fixed $\sigma$, this ansatz describes a non-commutative two-sphere with a physical radius given by $$R(\sigma)^2 = \frac{\lambda^2}{N_m}\sum\limits_{i=1}^{3}Tr[\phi_i (\sigma)^2] =\lambda^2 C\hat{R}(\sigma)^2,$$ with $C$ is the quadratic Casimir of the particular representation of the generators $\al_i$ defined by the identity $$\sum\limits_{i=1}^{3}(\al_i)^2=CI_{N_m}$$ where $I_{N_m}$ is the $N_m\times N_m$ identity matrix and $C=(N_m)^2-1$ for the irreducible $I_{N_m}$ is the $N_m\times N_m$ representation.

By computing the determinant, the action (7) becomes \beq S=-T_1\int d^2\sigma STr \Big[ (1-\lambda^2
E^2 + \al_i \al_i \hat{R}'^2)(1+4\lambda^2 \al_j \al_j \hat{R}^4 )\Big]^{1\over2}.\eeq
Hence, we get the funnel solution for dyonic string by solving the equation of motion of $\hat{R}$ as follows \beq \phi_i =\frac{\al_i}{2\sigma\sqrt{1-\lambda^2 E^2}}.\eeq

Now, we see if the two descriptions discussed above match or not.
As we showed in \cite{brokdual}, the two descriptions D1 and D3
don't have a complete agreement in the presence of a world volume
electric field since the energies don't match as we will see in the
following, even if their profiles match very well.

The energy is easily derived from the action (9) for the static solution (10). The minimum energy condition is $$\frac{d\phi_i}{d\sigma}=\pm\frac{i}{2}\epsilon_{ijk}[\phi_j,\phi_k],$$ which can be identified as the Nahm equations \cite{nahm}. We insert the ansatz (8) and this implies $$\hat{R}'=\pm 2\sqrt{1-\lambda^2 E^2}\hat{R}^2.$$ Using this condition and evaluating the Hamiltonian, $\int d\sigma (DE-L)$, for the dyonic funnel solutions ($L$ is the Lagrangian), the energy is expressed as
$$E_1=T_1 \int d\sigma STr \Big[\frac{\lambda^2 E^2}{\sqrt{1-\lambda^2 E^2}}+\sqrt{1-\lambda^2 E^2}|1+4\lambda^2 \al_j \al_j \hat{R}^4 |\Big].$$ We can manipulate this result by introducing the physical radius $R=\lambda\sqrt{C}|\hat{R}|$ and using $T_1=4\pi^2\ell_s T_3$. It's also useful to consider the electric displacement $D$ $$D
\equiv \frac{1}{N_m}\frac{\de S}{\de E}=\frac{\lambda^2 T_1 E}{\sqrt{1-\lambda^2 E^2}}=\frac{N_e}{N_m}.$$
Consequently, the energy from the D1 brane theory is found to be \beq E_1=T_1 \int d\sigma\sqrt{N_m^2
+g_s N_e^2}+T_3 (1-\frac{1}{N_m^2})^{\frac{-1}{2}}\int dR 4\pi R^2 ,
\eeq 
with $g_s$ is the string coupling with $T_1=(\lambda g_s)^{-1}$. The first term comes from collecting the contributions independent of $\hat{R}$. The second term gotten from the terms containing $\hat{R}$ and is used to put these in the form $\hat{R}^2|\hat{R}'|$. Then we have repeatedly applied $\hat{R}'=\pm 2\sqrt{1-\lambda^2 E^2}\hat{R}^2$ in producing the second term.

If we consider large $N_m$ the energy is reduced to the
following \beq E_1=T_1 N_m \int d\sigma, \eeq which can be rewritten
in terms of physical radius $R$ as \beq E_1=T_3 N_m \int 4\pi R^2 dR,
\eeq with $T_3 =\frac{T_1}{4\pi^2 \ell_s^2}$. In D3-brane
description the energy (3) becomes \beq E_3=T_3 \int
d^3\sigma\sqrt{1+\lambda^4 \frac{N_m^2 [(N_m +N_e )^2 +N_e^2]}{16r^8
} + 2\lambda^2 \frac{N_m (N_m +N_e )}{4r^4 }+2\lambda^2 \frac{ N_e^2
}{4r^4 }},\eeq such that the magnetic and the electric fields are
given by \beq \stackrel{\rightarrow}{B}=\frac{N_m
}{2r^2}\stackrel{\rightarrow}{r},
\phantom{~~~~}\stackrel{\rightarrow}{E}=\frac{N_e
}{2r^2}\stackrel{\rightarrow}{r}.\eeq In the large $N_m$ limit and
fixed $N_e$, the energy (14) of the spherically symmetric BPS
configuration is reduced to \beq E_3=T_3 \frac{N_m \sqrt{(N_m +N_e
)^2 +N_e^2}}{N_m +N_e} \int 4\pi r^2 dr. \eeq Again we consider large
$N_m$ limit and fixed $N_e$ and we get
$$\frac{ \sqrt{(N_m +N_e )^2 +N_e^2}}{N_m +N_e}\longrightarrow 1.$$
Consequently, for fixed $N_e$ and large $N_m$ limit we have
agreement from both sides (D1 and D3 descriptions) and the energy is
\beq E_3=T_3 N_m \int 4\pi r^2 dr, \eeq in which we identify the
physical radius $R$ from D1 description and $r$ from D3 description.

Now, if we take large $N_m$ limit keeping $N_e/N_m=K$ fixed at any
arbitrary $K > 0$ the result will be different. Thus, from D1
description the energy becomes \beq E_1=T_3 N_m \sqrt{1+g_s K^2}
\int 4\pi R^2 dR, \eeq and from D3 description the energy is \beq
E_3=T_3 \frac{N_m \sqrt{(1+K)^2 +K^2}}{1+K} \int 4\pi R^2 dR . \eeq
Then we have disagreement. Consequently, the presence of electric
field spoils the duality between D1 and D3 descriptions of
intersecting D1-D3 branes.
\section{Intersecting D1-D5 Branes at the Presence of an Electric Field}
\hspace{.3in}Although D1$\bot$D5
branes system \cite{f1} is not supersymmetric, the fuzzy funnel
configuration in which the D-strings expand into orthogonal
D5-branes shares many common features with the D3-brane funnel.
Thus, we are interested in establishing whether a similar result
holds also in the case of D1$\bot$D5 branes meaning the presence of
a world volume electric field leads to broken duality or not.

From the D1 description the system is described by the action (6). We
consider static configurations involving five (rather than three)
nontrivial scalars, $\phi_i$ with $i=1,...,5$. The proposed ansatz
for the funnel solution is \beq \phi_i (\sigma)=\pm
\hat{R}(\sigma)G_i ,\eeq with $\hat{R}(\sigma)$ is the (positive)
radial profile and $G_i$ are the matrices constructed in \cite{f2}
to provide a fuzzy four-sphere and to construct the string funnel
\cite{f1}. Also, $G_i$ are given by the totally symmetric $n$-fold
tensor product of $4\times4$ gamma matrices, and that the dimension
of the matrices is related to the integer $n$ by \beq N=\frac{(n +
1)(n + 2)(n + 3)}{6}. \eeq Then the solution is a fuzzy 4-sphere
with the physical radius \beq R(\sigma)=\lambda\sqrt{\frac{Tr(\phi_i
\phi_i)}{N}}=\sqrt{c}\lambda\hat{R}(\sigma), \eeq $c$ is the
"Casimir" associated with the $G_i$ matrices, i.e., $G_i G_i = c
1_N$, given by \beq c = n(n + 4). \eeq By inserting the ansatz into
(6) the action becomes \beq S = -NT1 \int d^2 \sigma
\sqrt{1-\lambda^2 E^2+ (R')^2}\Big[ 1 +
\frac{4R^4}{c\lambda^2}\Big], \eeq where the prime indicates the
derivative with respect to $\sigma$.

The electric field is fixed by the quantization condition on the
displacement field, $D=\frac{N_f}{N}$, where \beq
D=\frac{1}{N}\frac{\delta S}{\delta E}=\frac{\lambda^2 T_1
E}{\sqrt{1-\lambda^2 E^2}} \eeq after using the equations of motion,
the energy ($\tilde{E}_1 =\int d\sigma (DE-L)$) of the system is
evaluated to be \beq \tilde{E}_1 =\sqrt{N^2 +g_s^2 N_f^2}T_1 \int
d\sigma +\frac{6N}{c}T_5 \int \Omega_4 R^4 dR +NT_1 \int dR +\Delta
E, \eeq with $T_5 =\frac{T_1}{(2\pi \ell_s)^4}$ and the first and
the second terms correspond to the energies of N semi-infinite
strings stretching from $\sigma = 0$ to infinity and of $6N/c$
D5-branes respectively. The contribution of the last terms to
the energy indicates that the configuration is not supersymmetric.
The last contribution is a finite binding energy
$\Delta E =  1.0102Nc^{1/4}T_1 \ell_s$.

In the dual point of view of the D5-brane world volume theory, the
action describing the system is the Born-Infeld action \beq
S=-T_5\int d^6\sigma STr \sqrt{-det(G_{ab}+\lambda^2 \partial_a \phi
\partial_b \phi +\lambda F_{ab})} \eeq with $a,b=0,1,...,5$ and
$\phi$ the excited transverse scalar. To get a spike solution with
electric field switched on from D5-brane theory we follow the
analogous method of bion spike in D3-brane theory as discussed in
\cite{f1,f2}. We use spherical polar coordinates and the metric is
\beq ds^2 =G_{ab}d\sigma^a d\sigma^b =-dt^2 +dr^2 +r^2 g_{ij}d
\alpha^i d \alpha^j , \eeq with $r$ the radius and $\alpha^i$,
$i=1,...,4$, Euler angles. $g_{ij}$ is the diagonal metric on a
four-sphere with unit radius \beq g_{ij}=\pmatrix{1&&&\cr&sin^2
(\alpha^1)&&\cr&&sin^2 (\alpha^1)sin^2 (\alpha^2)&\cr&&&sin^2
(\alpha^1)sin^2 (\alpha^2)sin^2 (\alpha^3)\cr}. \eeq In this theory,
we add the electric field $E$ as a static radial field in the $U(1)$
sector. The scalar $\phi$ is only a function of the radius by
considering bion spike solutions with a "nearly spherically
symmetric" ansatz. In the same time to compare with the radial
profile obtained in D1-brane theory we identify the physical
transverse distance as $\sigma = \lambda\phi$, and the radius $r =
R$ which fixes the coefficients. Thus the scalar is found to be \beq
\phi(r)=\pm\int \frac{dr}{\lambda(1-\alpha^2)\Big(
(\frac{r^4}{\lambda^2 }+1)^2 -1\Big)} \eeq where $\alpha= \frac{g_s
N_f}{\sqrt{N^2 +g_s^2 N_f^2}}$. The energy of the system is
evaluated to be \beq \tilde{E}_5 =\frac{NT_1 }{\sqrt{1-\Big(
\frac{g_s N_f}{\sqrt{N^2 +g_s^2 N_f^2}}\Big)^2 }}\int d\sigma
+\frac{6N}{c}T_5 \int \Omega_4 R^4 dR +NT_1 \int dR +\Delta E, \eeq
with $\Delta E$ is the same one found above from D1 description. In
the absence of electric field, it's clear that by identifying the
profiles of D1 and D5 descriptions in the limit of $N$ we could get
complete agreement for the geometry and the energy determined by the
two dual approaches. Now, in the presence of an electric field it
seems there is also agreement. We compare the energy from the D1 description (26) and the energy from the D5 description (31). If we consider the large $N$ limit and
fixed $N_f$ the first term of $\tilde{E}_1$ becomes \beq \sqrt{N^2
+g_s^2 N_f^2}T_1 \int d\sigma \longrightarrow NT_1 \int d\sigma ,\eeq
and the first term of $\tilde{E}_5$ goes to the following value \beq
\frac{NT_1 }{\sqrt{1-\Big( \frac{g_s N_f}{\sqrt{N^2 +g_s^2
N_f^2}}\Big)^2 }}\int d\sigma \longrightarrow NT_1 \int d\sigma ,\eeq
which proves the agreement at large $N$.\\

Now, let's fix the value $\frac{g_s N_f}{N}$ to be one value $M$ which
can't be neglected at large limit of $N$. Thus, if $N$ is large the
last two limits (32) and (33) become \beq \sqrt{N^2 +g_s^2 N_f^2}T_1
\int d\sigma \longrightarrow NT_1 \sqrt{1+M^2}\int d\sigma \eeq and
\beq \frac{NT_1 }{\sqrt{1-\Big( \frac{g_s N_f}{\sqrt{N^2 +g_s^2
N_f^2}}\Big)^2 }}\int d\sigma \longrightarrow \frac{NT_1
}{\sqrt{1-\frac{M^2}{1+M^2}}}\int d\sigma.\eeq 
The right hand term of Eq(35) is equal to the right hand term of Eq(34) $\frac{NT_1}{\frac{1}{\sqrt{1+M^2}}\sqrt{1+M^2-M^2}}=NT_1 \sqrt{1+M^2}$.
Then, this implies agreement of the two duals at the
level of energy of the two descriptions. Consequently, the duality in
D1$\bot$D5 branes is unbroken by switching on the electric field.

\section{Two and Three Excited Scalars}
\hspace{.3in}In this section, we use the abelian Born-Infeld action
for the world volume gauge field and transverse displacement scalars
to explore some aspects of D3-brane structure and dynamics. We deal
with magnetic and dyonic cases.
\subsection{Absence of Electric Field}
\hspace{.3in}We consider the case where D3-brane has more than one scalar
describing transverse fluctuations. We denote the world volume
coordinates by $\sigma^a$, $a=0,1,2,3$, and the transverse
directions by the scalars $\phi^i$, $i=4,...,9$. In D3-brane theory
construction, the low energy dynamics of a single D3-brane is
described by the BI action by using static gauge \beq S_{BI}=\int
L=-T_3 \int d^4\sigma\sqrt{-det(\eta_{ab}+\lambda^2\partial_a \phi^i
\partial_b \phi^i +\lambda F_{ab})} \eeq with $F_{ab}$ is the field
strength of the $U(1)$ gauge field on the brane. By
exciting two scalar fields and setting to zero the other scalars, the energy is evaluated for the fluctuations through two directions and for static configurations as follows

\beq\bra{ll} \zeta= -L&=T_3 \int d^3\sigma\Big[ 1 +\lambda^2 \Big(
\mid \nabla\phi_4 \mid^2 +\mid \nabla\phi_5 \mid^2
+\stackrel{\rightarrow}{B}^2 \Big)+ \lambda^4
\Big( (\stackrel{\rightarrow}{B}.\nabla\phi_4 )^2+(\stackrel{\rightarrow}{B}.\nabla\phi_5 )^2 \Big)\\\\
&+\lambda^4 \mid\nabla\phi_4 \wedge\nabla\phi_5 \mid^2 \Big]
^{\frac{1}{2}}. \era\eeq

If we introduce a complex scalar field $C=\phi_4 + i\phi_5 $, we
can rewrite the energy as \beq\bra{ll} \zeta&= T_3 \int
d^3\sigma\Big[ \lambda^2 (\nabla C+\stackrel{\rightarrow}{B})(\nabla
C+\stackrel{\rightarrow}{B})^{*}+(1-\lambda^2 (\nabla
C.\stackrel{\rightarrow}{B}) )(1-\lambda^2 (\nabla
C.\stackrel{\rightarrow}{B}))^{*}\\\\&+\frac{1}{4}\lambda^4 \mid
\nabla C \wedge\nabla C^{*} \mid^2\Big]^{\frac{1}{2}}. \era\eeq In
this case, we observe that to get minimum energy
we can set the first term to zero and this will lead to \beq\nabla C
=- \stackrel{\rightarrow}{B}=\nabla \phi_4 +i \nabla \phi_5 .\eeq We
know that $\stackrel{\rightarrow}{B}$ is real, then $\nabla \phi_5
=0$ and thus in the "bi-excited scalar" system the lowest energy is
identified to D3$\bot$D1 system. This suggests that to study
the minimum energy configuration of the D3-brane system it is only
worthwhile to excite just one scalar.

Now, requiring $\nabla \phi_5 \neq0$ we get different energy bound.
The energy (38) can be rewritten as follows \beq\bra{ll}
\zeta&=T_3 \int d^3\sigma\Big[ \lambda^2  \mid \nabla\phi_4
+\nabla\phi_5 \mp \stackrel{\rightarrow}{B} \mid^2 + (1\pm \lambda^2
\stackrel{\rightarrow}{B}.(\nabla\phi_4 +\nabla\phi_5))^2 \\\\
&\pm 2\lambda^2 \nabla\phi_4 .\nabla\phi_5 +\lambda^4 \mid
\nabla\phi_4 \wedge\nabla\phi_5 \mid^2\Big]^{\frac{1}{2}}. \era\eeq
The new bound is now found with the following constraint
\beq\nabla\phi_4+\nabla\phi_5 = \pm \stackrel{\rightarrow}{B},\eeq
and $\pm 2\lambda^2 \nabla\phi_4 .\nabla\phi_5 \geq 0$ should also be
satisfied. Then the energy is \beq \tilde{\zeta}= T_3 \int
d^3\sigma\Big[ (1\pm \lambda^2
\stackrel{\rightarrow}{B}.(\nabla\phi_4 +\nabla\phi_5 ) )^2 +
\lambda^4 \mid \nabla\phi_4 \wedge\nabla\phi_5 \mid^2 \pm 2\lambda^2
\nabla\phi_4 .\nabla\phi_5\Big]^{\frac{1}{2}}. \eeq We choose without loss of generality $\nabla\phi_4
\perp\nabla\phi_5$.\\

Thus we get $\nabla^2 \phi_4 +\nabla^2 \phi_5 =0$ (by using the
Bianchi identity), and the solution is
\beq\phi_{4}+\phi_{5}=\pm\frac{N_m}{2r}.\eeq

This solution could be generalized by considering three excited
scalars. The energy is then found to be \beq\bra{ll} \zeta_3 &= T_3 \int
d^3\sigma\Big[ 1+\lambda^2 \Big( \sum\limits_{i=4}^{6} \mid \nabla
\phi_i \mid^2 +\stackrel{\rightarrow}{B}^2 \Big)+\lambda^4 \Big(
\sum\limits_{i=4}^{6} \mid \stackrel{\rightarrow}{B}.\nabla \phi_i
\mid^2+\frac{1}{2} \sum\limits_{i,j=4}^{6}
\mid \nabla \phi_i\wedge \nabla \phi_j \mid^2 \Big) \Big]^{\frac{1}{2}}\\\\
&\geq  T_3 \int d^3\sigma \Big[ (1\pm \lambda^2
\stackrel{\rightarrow}{B}.\sum\limits_{i=4}^{6}\nabla \phi_i )^2
+\frac{1}{2} \sum\limits_{i,j=4}^{6} \mid \nabla \phi_i\wedge \nabla
\phi_j \mid^2 \pm\frac{\lambda^2}{2}
\sum\limits_{i,j=4}^{6}\nabla\phi_i .\nabla\phi_j
\Big]^{\frac{1}{2}}.\era\eeq We should also consider $\pm\lambda^2
\sum\limits_{i,j=4}^{6}\nabla\phi_i .\nabla\phi_j\geq 0$ to get the lowest
energy configuration. This should be found by canceling
some of the terms in the second line of the expression given in
(44). The simplest way is to require the orthogonality of $\nabla
\phi_i$ and $\nabla\phi_j$. Then the lowest energy in the static gauge
is \beq \tilde{\zeta}_3 = T_3 \int d^3\sigma \Big[ (1\pm \lambda^2
\stackrel{\rightarrow}{B}.\sum\limits_{i=4}^{6}\nabla \phi_i )^2
+\frac{1}{2} \sum\limits_{i,j=4}^{6} \mid \nabla \phi_i\wedge \nabla
\phi_j \mid^2 \Big]^{\frac{1}{2}}, \eeq with the constraints
\beq\sum\limits_{i=4}^{6}\nabla \phi_i =\pm
\stackrel{\rightarrow}{B},\eeq and the solution is similar to the
"bi-excited system", we find \beq\sum\limits_{i=4}^{6} \phi_i =\pm
\frac{N_m}{2r}.\eeq

The solution obtained for each excited scalar field has one collective
coordinate in the D3-brane world volume theory. This is the
direction along which the brane extends away from the D3-brane.
Thus, this collective coordinate represents a "ridge" solution in
the D3-brane theory.

Now, we will look at another case in which the electric field is
present, and to see how the energy of the system could be minimized
and what kind of solutions we could obtain.
\subsection{Addition of an Electric Field}
\hspace{.3in}First we start by exciting two transverse directions
($\phi_4$ and $\phi_5$) with the electric field
$\stackrel{\rightarrow}{E}$ switched on. We consider as previous that
$\nabla\phi_4\bot\nabla\phi_5$. Then the energy of our system is
\beq\bra{lll} E_3&= T_3 \int d^3\sigma\Big[ 1 +\lambda^2 \Big( \mid
\nabla\phi_4 \mid^2 + \mid \nabla\phi_5 \mid^2 +\stackrel{\rightarrow}{B}^2 +\stackrel{\rightarrow}{E}^2
\Big)\\\\
&+\lambda^4 \Big[ (\stackrel{\rightarrow}{B}.\nabla\phi_4
)^2+(\stackrel{\rightarrow}{B}.\nabla\phi_5 )^2
+(\stackrel{\rightarrow}{E}.\stackrel{\rightarrow}{B})^2 +\mid
\nabla\phi_4 \wedge\nabla\phi_5\mid^2 +\mid
\stackrel{\rightarrow}{E}
\wedge\nabla\phi_4\mid^2 +\mid \stackrel{\rightarrow}{E} \wedge\nabla\phi_5\mid^2 \Big] \\\\
&+\lambda^6 \mid \stackrel{\rightarrow}{E}.(\nabla\phi_4
\wedge\nabla\phi_5 )\mid^2\Big]^{\frac{1}{2}}\\\\ &=T_3 \int
d^3\sigma\Big( \lambda^2 \mid \nabla\phi_4 + \nabla\phi_5
\pm(\stackrel{\rightarrow}{B}+\stackrel{\rightarrow}{E})\mid^2
+[1\mp\lambda^2 (\nabla\phi_4 +
\nabla\phi_5).(\stackrel{\rightarrow}{B}+\stackrel{\rightarrow}{E})]^2 \\\\
&- 2\lambda^2 B.E +\lambda^4(\stackrel{\rightarrow}{E}.\stackrel{\rightarrow}{B})^2 \\\\
&+ \lambda^4 \Big[ \mid \nabla\phi_4 \wedge\nabla\phi_5\mid^2 +\mid
\stackrel{\rightarrow}{E} \wedge\nabla\phi_4\mid^2 +\mid
\stackrel{\rightarrow}{E} \wedge\nabla\phi_5\mid^2
-(\stackrel{\rightarrow}{E}.\nabla\phi_4)^2
-(\stackrel{\rightarrow}{E}.\nabla\phi_5)^2\\\\
&-2\Big( \nabla\phi_4
.(\stackrel{\rightarrow}{E}+\stackrel{\rightarrow}{B})\nabla\phi_5
.(\stackrel{\rightarrow}{E}+\stackrel{\rightarrow}{B})+(\nabla\phi_4
.\stackrel{\rightarrow}{B})(\nabla\phi_4
.\stackrel{\rightarrow}{E})+ (\nabla\phi_5
.\stackrel{\rightarrow}{B})(\nabla\phi_5
.\stackrel{\rightarrow}{E})\Big) \Big]
\\\\ &+\lambda^6 \mid \stackrel{\rightarrow}{E}.(\nabla\phi_4 \wedge\nabla\phi_5
)\mid^2\Big)^{\frac{1}{2}}. \era\eeq By taking into consideration the
previous analysis in the absence of an electric field, the energy
$E_3$ becomes \beq\bra{lll} E_3&\geq T_3 \int d^3\sigma\Big(
[1\pm\lambda^2 (\nabla\phi_4 +
\nabla\phi_5).(\stackrel{\rightarrow}{B}+\stackrel{\rightarrow}{E})]^2
- 2\lambda^2 \stackrel{\rightarrow}{B}.\stackrel{\rightarrow}{E}
+\lambda^4(\stackrel{\rightarrow}{E}.\stackrel{\rightarrow}{B})^2\\\\
&+ \lambda^4 \Big[ \stackrel{\rightarrow}{E}^2
\mid\nabla\phi_4\mid^2 + \stackrel{\rightarrow}{E}^2
\mid\nabla\phi_5\mid^2 \Big] +\lambda^6 \mid
\stackrel{\rightarrow}{E}.(\nabla\phi_4 \wedge\nabla\phi_5
)\mid^2\Big)^{\frac{1}{2}}\era,\eeq This expression is consistent
with the fact that \beq\nabla\phi_4 + \nabla\phi_5
\pm(\stackrel{\rightarrow}{B}+\stackrel{\rightarrow}{E})=0\eeq and
\beq- 2\lambda^2 \stackrel{\rightarrow}{B}.\stackrel{\rightarrow}{E}
+\lambda^4(\stackrel{\rightarrow}{E}.\stackrel{\rightarrow}{B})^2\geq
0.\eeq We also used the following expression \beq\mid
\stackrel{\rightarrow}{E} \wedge\nabla\phi_5\mid^2
=\stackrel{\rightarrow}{E}^2 \mid \nabla\phi_5 \mid^2
-(\stackrel{\rightarrow}{E}.\nabla\phi_5)^2 .\eeq Then to get the
lowest energy in the presence of an electric field we require $-
2\lambda^2 \stackrel{\rightarrow}{B}.\stackrel{\rightarrow}{E}
+\lambda^4(\stackrel{\rightarrow}{E}.\stackrel{\rightarrow}{B})^2=
0$; i.e. $\stackrel{\rightarrow}{E}.\stackrel{\rightarrow}{B} =
\frac{2}{\lambda^2}$ or $\stackrel{\rightarrow}{E}\bot
\stackrel{\rightarrow}{B}$. With these simplifications, the energy
becomes \beq\bra{llll} \tilde{E}_3&= T_3 \int d^3\sigma\Big( \Big[
1\pm\lambda^2 (\nabla\phi_4 +
\nabla\phi_5).(\stackrel{\rightarrow}{B}+\stackrel{\rightarrow}{E})\Big]^2
+ \lambda^4 \stackrel{\rightarrow}{E}^2\Big[ \mid\nabla\phi_4\mid^2
+ \mid\nabla\phi_5\mid^2 \Big]\\\\&+\lambda^6 \mid
\stackrel{\rightarrow}{E}.(\nabla\phi_4 \wedge\nabla\phi_5
)\mid^2\Big)^{\frac{1}{2}}.\era\eeq 
In the following we consider $\stackrel{\rightarrow}{E}.\stackrel{\rightarrow}{B} =
\frac{2}{\lambda^2}$. We remark that the energy of D2$\bot$D3 brane is increased by the presence of the
electric field and by switching it off we obtain the lowest energy configuration obtained previously.

By solving (50) in the static gauge, using the Bianchi identity, we obtain the solution \beq\phi_4 +
\phi_5 = \mp\frac{ N_m +N_e}{2r},\eeq with $r^2 =\sum\limits_{a=1}^{3}(\sigma^a)^2$, $N_m$ and $N_e$ the magnetic and the electric charges respectively.

Now, by exciting three scalars, the energy is generalized to the
following \beq\bra{ll}
\xi_3&= T_3 \int d^3\sigma\Big[ 1+\lambda^2 ( \sum\limits_{i=4}^{6} \mid \nabla \phi_i \mid^2
+\stackrel{\rightarrow}{B}^2 +\stackrel{\rightarrow}{E}^2)\\\\
&+\lambda^4 \Big(
(\stackrel{\rightarrow}{E}.\stackrel{\rightarrow}{B})^2
+\sum\limits_{i=4}^{6} \mid \stackrel{\rightarrow}{B}.\nabla \phi_i
\mid^2 +\sum\limits_{i=4}^{6} \mid
\stackrel{\rightarrow}{E}\wedge\nabla \phi_i \mid^2 +
\frac{1}{2}\sum\limits_{i,j=4}^{6}
\mid \nabla \phi_i\wedge \nabla \phi_j \mid^2  \Big)\\\\
&+\lambda^6  \frac{1}{2}\mid
\stackrel{\rightarrow}{E}.(\sum\limits_{i,j=4}^{6}\nabla\phi_i
\wedge\nabla\phi_j )\mid^2 \Big]^{\frac{1}{2}} \era\eeq

We also consider as before $\nabla\Phi_i \bot\nabla\Phi_j$ and
$\stackrel{\rightarrow}{E}.\stackrel{\rightarrow}{B}=\frac{2}{\lambda^2}$
with $i,j=4,5,6$. Then the energy will be reduced to the lowest energy for
three excited directions \beq\bra{ll} \tilde{\xi}_3&= T_3 \int
d^3\sigma\Big( [1\pm\lambda^2 \sum\limits_{i=4}^{6}\nabla \phi_i
.(\stackrel{\rightarrow}{B}+\stackrel{\rightarrow}{E})]^2 +
\lambda^4 \sum\limits_{i,j=4}^{p}\stackrel{\rightarrow}{E}^2 ( \mid
\nabla \phi_i \mid^2
+ \mid \nabla\phi_j  \mid^2 )\\\\
&+\lambda^6 \mid
\stackrel{\rightarrow}{E}.(\sum\limits_{i,j=4}^{6}\nabla\phi_i
\wedge\nabla\phi_j )\mid^2\Big)^{\frac{1}{2}} ,\era\eeq with $i<j$ in
the summations and where we assume the following condition
\beq\sum\limits_{i=4}^{6}\nabla \phi_i \pm
(\stackrel{\rightarrow}{B}+\stackrel{\rightarrow}{E})=0.\eeq 
The last equation is easily solved to give the solution
\beq\sum\limits_{i=4}^{6} \phi_i =\mp \frac{ N_m + N_e}{2r}.\eeq
Again if we require that $\stackrel{\rightarrow}{E}$ is parallel to
one of $\nabla\phi_i$ the energy will be minimized
as the last term in the expression (56) of the energy vanishes. Then,
accordingly to (56) and (58) with
$$|\stackrel{\rightarrow}{B}|=\left|\frac{N_m}{2r^2}\right|,
\phantom{~~~~}|\stackrel{\rightarrow}{E}|=\left|\frac{N_e}{2r^2}\right|$$
the energy becomes \beq\bra{lll} \tilde{\xi}_3= T_3 \int
d^3\sigma\Big( [1+\lambda^2 \frac{(N_m +N_e)^2}{4r^4}]^2 + \lambda^4
\frac{N_e^2 (N_m +N_e)^2}{8r^6} \Big)^{\frac{1}{2}} \era\eeq This is
the lowest energy for both cases where two or three transverse
scalars are excited since the solution is a superposition
of the scalars.
\section{Discussion and Conclusion}
\hspace{.3in}In this work we investigated the physics of intersecting D1-D5 branes in the presence of an electric field. We also studied D2$\bot$D3 branes and D3$\bot$D3 branes in presence/absence of an electric field.

Our main interest was the fate of the duality of D1$\bot$D5 branes. We found that the duality between the D1 and D5 descriptions is unbroken in the presence of an electric field. Then, the duality in D1-D5 case is still valid. By contrary, as we saw in \cite{brokdual}, the duality in D1$\bot$D3 branes is no longer valid when we add an electric field. Then we have further argued that the D1-brane description, in D1-D3 case, breaks down. This is strengthened by the result discussed in \cite{tensionstring} which have argued that the effective tension of the string  goes to zero. Thus, excited strings modes will not be very heavy compared to massless string modes and one might question the validity of the Dirac-Born-Infeld action which retains only the massless modes.

The investigation of excited D3-branes through the two cases of absence or presence of an electric field lead to the fact that by exciting 2 and 3 of its transverse directions the brane develops a spike which is interpreted as an attached bundle of a superposition of coordinates of another brane given as a collective coordinate a long which the brane extends away from the D3-brane. In our study of D3-branes, by exciting 2 and 3 transverse directions we found that a magnetic monopole produces a singularity in the D3-branes transverse displacement which can be interpreted as a superposition of coordinates describing D$p$-branes ($p=2,3$) attached to the D3-brane and the same for the dyonic case. We also obtained another important result that the lowest energy in the intersection branes case is obtained at the level of D1$\bot$D3 branes and the energy is higher if we excite more scalar fields and even more so in the presence of an electric field.
\section{Acknowledgments}

\hspace{.3in}This work was supported by a grant from the Arab Fund
for Economic and Social Development. The author would like to thank
R. Bhattacharyya for very pleasant and helpful discussion at an early stage of this work, Robert de Mello Koch for interesting remarks and Robert Myers for his important remark concerning the duality. Also special thanks to Stellenbosch Institute for Advanced Study where a part of this work was started.

\end{document}